\documentclass[aps,prd,nofootinbib]{revtex4}

\usepackage{graphicx}
\usepackage{float}

\usepackage{epsfig}
\usepackage{color}

\newcommand{\nn}{\nonumber \\}

\usepackage{amsmath}

\usepackage{amssymb}

\begin{document}
\title{ Cosmological Solutions in $f(Q)$  Gravity via  Noether Symmetry Approach}
\author{ M. Mahmoudzadeh Baghbani}\email{mahmoudzadeh67@gmail.com}
\author{K. Atazadeh}\email{atazadeh@azaruniv.ac.ir}
\author{M. Mousavi}\email{mousavi@azaruniv.ac.ir}
\affiliation{Department of Physics, Azarbaijan Shahid Madani University, Tabriz, 53714-161 Iran}

\begin{abstract}
Symmetry plays a crucial role in theoretical physics, especially Noether symmetry, which is a powerful approach for identifying the models at the fundamental level. The exact solution is provided within the point-like Lagrangian framework. In this work, we study one of the alternative theories of gravity based on the non-metricity scalar $Q$, namely $f(Q)$ gravity, via Noether symmetry.
We utilize Noether symmetry within the framework of $f(Q)$ gravity to derive the functional expression for $f(Q)$, which is given by $f(Q)=c(Q-nQ)^{\frac{3}{2-2n}}$.
To confirm the exact solution of the model through Noether symmetry, we continue to consider the Friedmann-Robertson-Walker (FRW) cosmology with the dynamical solution of the system using dimensionless variables and show that the accelerated expansion of the universe follows a power law scale factor.
In the following, we show that the quantities corresponding to the exact solution for $n<1$ lead to an accelerated expansion universe.
Finally, in the framework of $f(Q)$ scalar-tensor cosmology, we apply the Noether symmetry approach to find the cosmological models consistent with the Noether symmetry.
\end{abstract}

\maketitle


\section{Introduction}
\label{section1}

The investigation and theoretical modeling of the accelerated expansion of the universe \cite {1,2,3} is one of the most active areas of research in cosmology during the last three decades. It's understood that an unknown energy component, so-called dark energy, or a generalization of general relativity, could be responsible for this phenomenon. Currently, various generalized gravitational theories are considered as competing alternatives, much like the dark energy scenario. Numerous studies have been carried out in recent decades, including studies on modified gravity models such as $f(R)$ theory and $f(T)$ theory,
which has attracted attention in this field \cite{4,5,6,7,77,777,7777}. It should be noted that in all these models, metric compatible manifolds have been used.
Another theory of gravity that we examine through the Noether symmetry approach is the theory of $f(Q)$ gravity \cite{8,9,10}.
In fact, a generalization of the symmetry teleparallel gravity \cite {11}, which has recently attracted interest, is the $f(Q)$ gravity theory.
In the $f(Q)$ theory, gravity via  $Q$ is coupled non-minimally to the matter Lagrangian. As a cosmological application of the $f(Q)$ gravity, it can represent an alternative approach to dark energy. One of the essential features of the $f(Q)$ theory is that, unlike general relativity, we can also separate gravity from the inertial effect. It is also worth mentioning that while the field equations in  $f(R)$ gravity are fourth-order \cite{13,133}, but they are of second-order in $f(Q)$ gravity, and hence, $f(Q)$ gravity is free from pathologies. Thus, the construction of this theory forms a novel starting point for various modified gravity theories \cite{14,15,155}. Various works in the literature suggest that the $f(Q)$ theory is one of the promising alternative formulations of gravity to explain cosmological observations \cite{16,17,18,19,20}. Especially in Ref. \cite{200}, the authors attempt to construct a convenient $f(Q)$ gravity by utilizing well-known cosmological solutions, such as inflationary and dark energy models. However, in this study, we apply the Noether symmetry approach in the context of $f(Q)$ gravity to obtain exact cosmological solutions. The advantage of using the Noether symmetry approach in gravitational actions, such as $f(Q)$ gravity, is to search for exact solutions in the model without requiring further assumptions.\\
$f(Q)$ gravity is based on non-metricity $Q$ scalar and the manifold in this theory is not metric compatible, that is, $Q_{\alpha\mu\nu}$=$\nabla_{\alpha}g_{\mu\nu}$, in fact, this model is more general than the previous models \cite{21,22}. In addition to that, in general relativity \cite{24,25}, the affine connection \cite{26,27} is used instead of the Levi-Civita connection, and one of the components of this connection is the disformation connection, which result from the parallel transport a vector field along the curve because in the parallel transport of a vector field, its magnitude changes, this change is measured for two points very close to each other by the non-metricity tensor $Q_{\alpha\mu\nu}$. In this paper, the Noether symmetry approach is applied to search for cosmological solutions in generic $f(Q)$ theories of gravity \cite{28,288,29,30}. We consider Noether symmetry in the FRW universe and search for an $f(Q)$ Lagrangian compatible with it.\\
The dynamical system approach provides a powerful framework to study the evolution of the universe by transforming the modified gravity field equations into a set of autonomous differential equations \cite{00}. Instead of solving highly complex equations directly, one analyzes the phase space of the system, where critical points correspond to cosmological epochs. This method not only simplifies the analysis of nonlinear dynamics but also offers a unified way to compare different cosmological models within general relativity and modified gravity theories. Here, we examine the stability of the fixed points in the framework of $f(Q)$ gravity which it obtained via Noether symmetry and we try to determine which solutions are physically viable for the long-term evolution of the universe. \\
We note that recently, during the preparation of this paper, a study appeared \cite{300} concerning Noether symmetry in $f(Q)$ gravity, which differs from the current work in both content and results.
\\
The plan of the paper is as follows. In section II, we study the Noether symmetry, in which the point-like Lagrangian plays an important role. In section III, we give a short review of the $f(Q)$ gravity. In section IV, the exact solution of the theory via the Noether symmetry approach shows that the universe experiences a power-law expansion. In section V, by using the dynamical system approach, we analyze the cosmological dynamics of the model, and we extract the critical points of the system by solving the equations $x'=0$ and $y'=0$. In section VI, in the framework of $f(Q)$ scalar-tensor cosmology, we apply the Noether symmetry approach to find the cosmological models consistent with the Noether symmetry.
Finally, we give a conclusion in section VII.


\section{A Brief Review of Noether's Symmetry}
\label{sec:noether}

In this section, we outline Noether's symmetry framework, a cornerstone for constructing gravitational theories consistent with space-time symmetry.

Consider a non-degenerate canonical Lagrangian \( L(q^{i}, \dot{q}^{i}) \), where \( q^{i} \) denotes generalized coordinates and \( \dot{q}^{i} = \frac{dq^{i}}{d\lambda} \) represents their derivatives with respect to an affine parameter \( \lambda \) (often corresponding to time \( t \)). Noether's approach imposes the condition
\begin{equation}\label{eq:lambda_independence}
  \frac{\partial L}{\partial \lambda} = 0,
\end{equation}
indicating the Lagrangian's independence from \( \lambda \). Furthermore, the non-degeneracy of \( L \) is ensured by the non-vanishing Hessian determinant
\begin{equation}\label{eq:hessian}
  \det H_{ij} = \det\left\Vert\frac{\partial^{2}L}{\partial\dot{q}^{i}\partial\dot{q}^{j}}\right\Vert \neq 0,
\end{equation}
where \( H_{ij} \) is the Hessian matrix associated with \( L \).

In analytical mechanics, the Lagrangian is conventionally expressed as the difference between kinetic energy \( T(q, \dot{q}) \) and potential energy \( V(q) \)
\begin{equation}\label{eq:lagrangian}
  L = T(q, \dot{q}) - V(q).
\end{equation}
The corresponding Hamiltonian (or total energy function) is derived via the Legendre transformation as
\begin{equation}\label{eq:hamiltonian}
  E = \left(\frac{\partial L}{\partial \dot{q}^{i}}\right) \dot{q}^{i} - L.
\end{equation}
Under smooth derivative transformations, any invertible and smooth coordinate transformation \( Q^i = Q^i(q) \) induces a corresponding velocity transformation
\begin{align}
\label{eq:velocity_transform}
\dot{Q}^j(q) &= \frac{\partial Q^i}{\partial q^j} \dot{q}^j.
\end{align}

If the coordinate transformation depends on a parameter \(\varepsilon\), such that \( Q^i = Q^i(q, \varepsilon) \), it forms a one-parameter Lie group. For infinitesimal transformations, this is generated by a vector field
\begin{align}
\label{eq:vector_field}
\mathbf{X} &= \alpha^i(q) \frac{\partial}{\partial q^i},
\end{align}
whose canonical lift (prolongation) to the tangent bundle is
\begin{align}
\label{eq:prolongation}
X^c &= \alpha^i(q) \frac{\partial}{\partial q^i} + \left( \frac{d}{d\lambda} \alpha^i(q) \right) \frac{\partial}{\partial \dot{q}^j}.
\end{align}
Here, \( X^c \) is generator of  Noether's symmetry \cite{31}. A function \( f(q, \dot{q}) \) is invariant under this transformation if:
\begin{align}
\label{eq:lie_derivative}
\mathcal{L}_X f &\equiv \alpha^i(q) \frac{\partial f}{\partial q^i} + \left( \frac{d}{d\lambda} \alpha^i(q) \right) \frac{\partial f}{\partial \dot{q}^j} = 0.
\end{align}
If \( \mathcal{L}_X L = 0 \), the Lagrangian \( L \) admits \( X^c \) as a symmetry.

For a Lagrangian \( L \) satisfying the Euler-Lagrange equations
\begin{align}
\label{eq:euler_lagrange}
\frac{d}{d\lambda} \left( \frac{\partial L}{\partial \dot{q}^i} \right) - \frac{\partial L}{\partial q^i} &= 0,
\end{align}
consider the vector field \( X^c \) from \eqref{eq:prolongation}. Contracting \eqref{eq:euler_lagrange} with \( \alpha^j \)
\begin{align}
\label{eq:contracted_EL}
\alpha^j \left[ \frac{d}{d\lambda} \left( \frac{\partial L}{\partial \dot{q}^j} \right) - \frac{\partial L}{\partial q^j} \right] &= 0.
\end{align}
Using the product rule
\begin{align}
\label{eq:product_rule}
\frac{d}{d\lambda} \left( \alpha^j \frac{\partial L}{\partial \dot{q}^j} \right) &= \alpha^j \frac{d}{d\lambda} \left( \frac{\partial L}{\partial \dot{q}^j} \right) + \frac{d\alpha^j}{d\lambda} \frac{\partial L}{\partial \dot{q}^j},
\end{align}
we rewrite \eqref{eq:contracted_EL} as
\begin{align}
\label{eq:noether_intermediate}
\frac{d}{d\lambda} \left( \alpha^j \frac{\partial L}{\partial \dot{q}^j} \right) &= \mathcal{L}_X L.
\end{align}
If \( \mathcal{L}_X L = 0 \), the quantity
\begin{align}
\label{eq:noether_charge}
Q_0 &= \alpha^j \frac{\partial L}{\partial \dot{q}^j} = \alpha \frac{\partial L}{\partial \dot{a}} + \beta \frac{\partial L}{\partial \dot{Q}} = \text{Const.},
\end{align}
is conserved.
This \( Q_0 \) is the Noether charge \cite{32,33}, corresponding to the symmetry generated by \( X^c \).

\section{$f(Q)$ gravity }
\label{section3}
In the framework of generalized $ f(Q) $ gravity, it is necessary to say that the $Q$ scalar and the  non-metricity tensor $Q_{\alpha\mu\nu}$ play a central role in characterizing deviations from standard Riemannian geometry \cite{301}. When a vector field $V$ is transported parallel along the curve, the magnitude of the vector field changes, thus this change is measured by the $Q_{\alpha\mu\nu}$ non-metricity tensor for two points $p$ and $q$ very close to each other as
\begin{eqnarray}\label{1}
 \parallel V_{q} \parallel^{2} - \parallel V_{p}\parallel^{2} = Q_{\alpha\mu\nu} (p) U_{p}^{\alpha} V_{p}^{\mu} V_{q}^{\nu}.
\end{eqnarray}
Equation (\ref{1}) allows us to understand how the magnitude of the vector field changes in parallel transport.
 Using Leibnitz's theorem we can write
\begin{eqnarray}\label{2}
 \frac{d}{d t} P(\gamma)^{t}_{0} \parallel V \parallel^{2} = U^{\alpha} \nabla_{\alpha} (g_{\mu\nu} V^{\mu}V^{\nu})= (U^{\alpha} \nabla_{\alpha}
g_{\mu\nu})V_{\mu}V^{\nu} + 2g_{\mu\nu} (U^{\alpha} \nabla_{\alpha} V^{\mu}) V^{\nu} = Q_{\alpha\mu\nu} V^{\mu} V^{\nu},
\end{eqnarray}
The second term in equation (\ref{2}) is eliminated in parallel transport.
By introducing the non-metricity condition as
\begin{eqnarray}\label{3}
\nabla_{\alpha} g_{\mu\nu} \neq0,
\end{eqnarray}
and the connection $\Gamma^{\alpha}_{\mu\nu}$ that it satisfies the this condition, is metric compatible connection.
It should be noted that $Q^{~~\mu\nu}_{\alpha}\neq \nabla_{\alpha} g^{\mu\nu}$, by calculating the
covariant derivative form $g_{\mu\lambda}g^{\lambda\nu}= \delta_{\mu}^{\nu}$, it can be easily shown that
\begin{eqnarray}
Q^{~~\mu\nu}_{\alpha}= -\nabla_\alpha g^{\mu\nu}.
\end{eqnarray}
The simplest geometric interpretation for the $Q_{\alpha\mu\nu}$ tensor,
the change in the magnitude of a vector field in parallel transport which it is measured by the non-metricity tensor.
In general, it can be said that in parallel transport, the non-metricity tensor measures how the quantities depend on the metric change.
For example, this tensor measures how the angle between two vectors change in parallel transmission.
For example, how to measure the volume of $n$-dimension a area which is transferred along the $\gamma$ curve parallel to the tangent vector $u$.
\begin{eqnarray}
{\rm Vol} = \int \sqrt{|g |} d^{n} x,
\end{eqnarray}
\begin{eqnarray}
\frac{d}{d t} {\rm Vol}(\Omega) = \int_{\Omega} u^{\alpha} (\nabla_{\alpha} \sqrt{|g|})d^{n} x = \frac{1}{2} \int_{\Omega} \sqrt{|g|} u^{\alpha}
g^{\mu\nu} \nabla_{\alpha} g_{\mu\nu} d^{n} x = \frac{1}{2} \int_{\Omega} \sqrt{|g |} u^{\alpha} g^{\mu\nu} Q_{\alpha\mu\nu} d^{n} x,
\end{eqnarray}
Due to the appearance of the non-metricity tensor trace $g^{\mu\nu}Q_{\alpha\mu\nu} $, it is better to introduce a symbol for this trace.
The non-metricity tensor has three indices and the last two are symmetric, we can introduce two independent trace
\begin{eqnarray}
 Q_{\alpha}=g^{\mu\nu} { Q_{\alpha\mu\nu}} =Q_{\alpha}~ ^{\mu}~ _{\mu},
\end{eqnarray}
\begin{eqnarray}
\tilde{Q}_{\alpha}=g^{\mu\nu} Q_{\mu\nu\alpha}=Q{^{\mu}  ~ _{\mu\alpha}}.
\end{eqnarray}
We must mention the geometry we will work on the geometry of symmetry teleparallel equivalent in general relativity (STEGR), in which the
non-metricity tensor, torsion tensor and Riemann tensor are as follows
\begin{eqnarray}
{Q_{\alpha\mu\nu} \neq0 , ~~~~~~         T^{\alpha}\, _{\mu\nu}\neq0  ,  ~~~~~~~           R^{\alpha} ~ _{\mu\nu\rho} =0 }.
\end{eqnarray}
In the general affine-metric geometry, our desired geometry includes manifold, metric and connection. Therefore, we can covariant derivative of
metric and the cyclic permutations of indices and finally reach the separation of the connection. Since the torsion tensor is the a non-symmetric part of the
affine connection, we should have studies on the  Levi-Civita connection and on the non-metricity tensor. Now we turn to the covariant derivative of the metric
\begin{eqnarray}
\nabla_{\alpha}g_{\mu\nu} = \partial_\alpha g_{\mu\nu} -\Gamma^{\beta}\, _{\alpha\nu}g_{\mu\beta} -\Gamma^{\beta}\,_{\alpha\mu} g_{\beta\mu},
\end{eqnarray}
\begin{eqnarray}
\nabla_{\mu} g_{\nu\alpha} = \partial _{\mu} g_{\nu\alpha} -\Gamma^{\beta}\,_{\mu\alpha} g_{\nu\beta}- \Gamma^{\beta}\,_{\mu\nu} g_{\beta\alpha},
\end{eqnarray}
\begin{eqnarray}
\nabla_{\nu} g_{\alpha\mu}=\partial_\nu g_{\alpha\mu} -\Gamma^{\beta}\,_{\nu\alpha}g_{\beta\mu} -\Gamma^{\beta}\,_{\nu\mu} g_{\alpha\beta},
\end{eqnarray}
After adding together the first two equation and subtracting the last one, we obtain
\begin{eqnarray}
Q_{\alpha\mu\nu} +Q_{\mu\nu\alpha} -Q_{\nu\alpha\mu} &=&\partial_\alpha g_{\mu\nu} +\partial_\mu g_{\nu\alpha}-\partial _{\nu} g_{\alpha\mu}
-\Gamma^{\beta}_{\alpha\nu}g_{\mu\beta} -\Gamma^{\beta}_{\mu\alpha} g_{\nu\beta} +\Gamma^{\beta}_{\nu\mu} g_{\alpha\beta}-\\\nonumber
&&\Gamma^{\beta}_{\alpha\mu} g_{\beta\nu} -\Gamma^{\beta} _{\mu\nu} g_{\beta\alpha}+\Gamma^{\beta}_{\nu\alpha}g_{\beta\mu},
\end{eqnarray}
\begin{eqnarray}
Q_{\alpha\mu\nu} +Q_{\mu\nu\alpha} -Q_{\nu\alpha\mu}=\partial \alpha g_{\mu\nu} +\partial \mu g_{\nu\alpha}-\partial _{\nu}
g_{\alpha\mu}+T^{\beta}_{\nu\mu}g_{\alpha\beta}+T^{\beta}_{\nu\alpha}g_{\mu\beta}+T^{\beta}_{\alpha\mu}g_{\nu\beta}-2\Gamma^{\beta}_{\alpha\mu}g_{\beta\nu},
\end{eqnarray}
Contracting with $\frac{1}{2}( g^{\beta\nu})$ we obtain,
\begin{eqnarray}
\Gamma^{\beta}~_{\alpha\mu}=\{^{\beta}~_{\alpha\mu}\}+K^{\beta}~_{\alpha\mu}+L^{\beta}~_{\alpha\mu},
\end{eqnarray}
this compact form of the decomposed connection we have introduced the contorsion tensor $K^{\beta}_{~~\alpha\mu}$ and the disformation
tensor $ L^{\beta}_{~~\alpha\mu}$,
\begin{eqnarray}
\{^{\beta}~_{\,\,\mu\nu}\} = \frac{1}{2}g^{\beta\lambda}(\partial_\nu g_{\mu\lambda}+\partial_\mu g_{\lambda\nu} -\partial _{\lambda}g_{\mu\nu}),
\end{eqnarray}
\begin{eqnarray}
K^{\beta}~_{~\mu\nu}=\frac{1}{2}T^{\beta} ~_{\mu\nu}+T^{~\beta}_{(\mu~~ \nu)},
\end{eqnarray}
\begin{eqnarray}
L^{\beta} ~_{~\mu\,\nu}=\frac{1}{2}Q^{\beta} ~_{\mu\,\nu} -Q^{~\beta}_{(\mu~\nu)},
\end{eqnarray}
In this coincident gauge the affine connection is zero, then the controsion will be zero and christofel's symbol
will be the correspondence of disformation connection.
\begin{eqnarray}
L^{\beta}_{\,\,\mu\nu}= - \{^{\beta}_{\,\,\mu\nu}\}.
\end{eqnarray}
The non-metricity tensor has two independent traces, the square non-metricity scalar can be defined as follows
\begin{eqnarray}
Q\equiv -\frac{1}{4}(Q_{\alpha\mu\nu}
Q^{\alpha\mu\nu})+\frac{1}{2}(Q_{\alpha\mu\nu}Q^{\mu\nu\alpha})+\frac{1}{4}Q_{\alpha}Q^{\alpha}-\frac{1}{2}Q^{\alpha}\tilde{Q^{\alpha}},
\end{eqnarray}
\begin{eqnarray}
\partial _{\alpha} g_{\mu\nu} -\Gamma^{\lambda}_{\alpha\mu}g_{\lambda\nu}-\Gamma^{\lambda}_{\alpha\nu}g_{\mu\lambda}=\nabla_{\alpha}g_{\mu\nu}.
\end{eqnarray}
Because in the coincident gauge the affine connection is zero, so we have $Q_{\alpha\mu\nu}=\partial_\alpha g_{\mu\nu}$.\\
In the flat FRW space-time the metric has the following form
\begin{eqnarray}
ds^{2}=-d t^{2}+a^{2}(t)\delta_{ij}d x^{i}d z^{j},
\end{eqnarray}
where $a(t)$ is scale factor of the universe.
For the above metric, $Q_{\alpha\mu\nu}=\partial_\alpha g_{\mu\nu}$ tensor has a non-zero components only for $\alpha=0$.
The non-zero components of the non-metricity tensor give
\begin{eqnarray}
~~~~~~~~~~~~~~ Q_{\alpha\mu\nu}Q^{\alpha\mu\nu}=-12H^{2}.
\end{eqnarray}
Also, for the flat FRW metric we obtain $Q=-6H^{2}$, where $ H\equiv\frac{\dot{a}}{a}$ is the Hubble parameter.

\section{$f(Q)$ cosmology via Noether symmetry}
\label{section4}
We consider the following action
\begin{eqnarray}
S=\int d^{4 } x\sqrt{-g} f(Q) +S_{m} ,
\end{eqnarray}
here we use the units $16\pi G =c=1$ and in the above action $S_{m}$ is the action of pressureless matter with minimally coupled with gravity.\\
In the FRW universe, one can define a canonical like-point Lagrangian,
$ L=L (a,\dot{a}, Q,\dot{Q})$, where $L$ is called Lagrangian density. The Lagrangian $L$ becomes canonical when selecting a suitable Lagrange multiplier and integrating by parts. In our  case, we have
\begin{eqnarray}\label{4.2}
S=2\pi^{2}\int d t a^{3}\left [f(Q)-\zeta\left( Q+ 6\frac{\dot{a}^{2}}{a^{2}}\right)-\frac{\rho_{m0}}{a^{3}}\right],
\end{eqnarray}
where $\rho=\rho_{m}= \frac{\rho_{m0}}{a^{3}}$ and $\zeta$ is a Lagrange multiplier,
the variation with respect to $Q$ of this action gives $\zeta=\frac{d f}{d Q}=f_{Q},$ and considering $f(Q)=Q+F(Q)$ the action (\ref{4.2}) is rewritten as follows
\begin{eqnarray}\label{4.3}
S=\int dt a^{3}\left [Q+F-\Big(1+F_{Q}\left(Q+6\frac{\dot{a}^{2}}{a^{2}}\right)\Big)-\frac{\rho_{m0}}{a^{3}} \right],
\end{eqnarray}
thus the like-point Lagrangian can be written as follows
\begin{eqnarray}\label{4.4}
L=a^{3}(F-QF_{Q})-6a\dot{a}^{2}-6a \dot{a}^{2} F_{Q}-\rho_{m0},
\end{eqnarray}
from equations (\ref{4.4}) and (\ref{eq:euler_lagrange}), one can find that
\begin{eqnarray}
\frac{\partial L}{\partial\dot{a}}=-12a\dot{a}(1+F_{Q}),
\end{eqnarray}
\begin{eqnarray}\label{4.5}
\frac{\partial L}{\partial a}=3a^{2}[F-QF_{Q}]-6F_{Q}\dot{a}^{2}-a^{3}QF_{QQ}-6F_{QQ}a\dot{a}^{2}-6\dot{a}^{2}.
\end{eqnarray}

The derivative of equation (\ref{4.5}) with respect to $t$ gives
\begin{eqnarray}\label{4.6}
\frac{d}{d t}\left(\frac{\partial L}{\partial \dot{a}}\right)=-12 \Big [F_{QQ}a\dot{a}\dot{Q}+(1+F_{Q})(\dot{a}^{2}+a\ddot{a}) \Big ],
\end{eqnarray}
Substituting above equations into the Euler-Lagrange equation (\ref{eq:euler_lagrange}) and using the spatially flat Friedmann equation in standard form as                                  \begin{eqnarray}\label{4.7}
-2 \left(\frac{\ddot{a}}{a}\right)-\frac{\dot{a}^{2}}{a^{2}}= { \frac{P}{2}},
\end{eqnarray}
 therefore, in the FRW space-time, we obtain
\begin{eqnarray}\label{4.8}
P=4\dot{H}(F_{Q}+2QF_{QQ})-\rho_{de},
\end{eqnarray}
where $\frac{\ddot{a}}{a}=\dot{H}+H^{2}$.
We assume that the matter is a prefect fluid whose $P$ and $\rho$ are the pressure and energy density of a prefect fluid, respectively and $\rho_{de}$ is the energy density of dark energy.
Here, we consider the case that the universe is filled with dust and radiation fluid, therefore we have the total pressure $P=P_{r}+P_{de}$ and the total energy density
$\rho=\rho_{m}+\rho_{r}+\rho_{de}$ in which $P_{m}=0$, $P_{r}=\frac{\rho_{r}}{3}$.
Where $P_{r}$, $P_{m}$ and $P_{de}$ are radiation pressure, matter pressureless and dark energy pressure, respectively.
From equation (\ref{4.8}) and using the relation $2\dot{H}+3H^{2}=-\frac{\rho_{r}}{3}-\rho_{de}$, we can write
\begin{eqnarray}\label{4.9}
P_{de}=4\dot{H}(F_{Q}+2QF_{QQ})-\rho_{de}-\frac{\rho_{r}}{3}.
\end{eqnarray}
By using equation (\ref{4.4}) and substituting it in the total energy (Hamiltonian) equation (\ref{eq:hamiltonian}), we have
\begin{eqnarray}\label{4.10}
E=a^{3}(-12H^{2}F_{Q}-F-6H^{2}-\rho_{m}).
\end{eqnarray}
Considering that the total energy is zero and substituting the non-metricity scalar in FRW metric by $Q=-6H^{2}$ in equation (\ref{4.10}), we can get
\begin{eqnarray}
2QF_{Q}-F+\rho_{m}=6H^{2}.
\end{eqnarray}
By taking
\begin{eqnarray}
\rho=\rho_{m}+\rho_{de},
\end{eqnarray}
and according to Friedmann equation in the spatially flat space-time as
\begin{eqnarray}
3H^{2}=8\pi\, G\rho,
\end{eqnarray}
 the dark energy density can be write as follows
\begin{eqnarray}
\rho_{de}= 2QF_{Q}-F.
\end{eqnarray}
Now, we use Noether symmetry in our model to use this approach to get the exact solution of the theory through the given Lagrangian.
As is well known, Noether symmetry is a useful tool for selecting models motivated at the fundamental level and finding the exact solution to the given Lagrangian.
In the following, the generator of the Noether symmetry vector in the configuration space is written as
\begin{eqnarray}
X=\alpha\frac{\partial}{\partial a}+\beta\frac{\partial}{\partial Q}+\dot{\alpha}\frac{\partial}{\partial \dot{a}}+\dot{\beta}\frac{\partial}{\partial \dot{Q}},
\end{eqnarray}
where $\alpha=\alpha(a,Q)$ and $\beta=\beta(a,Q)$ are both functions of the generalized coordinates $a$ and $Q$, Noether symmetry exists if the below equation is satisfied
\begin{eqnarray}\label{4.17}
{\cal L}_{X}L=XL=\alpha\frac{\partial L}{\partial a}+\beta\frac{\partial L}{\partial Q}+\dot{\alpha}\frac{\partial L}{\partial \dot{a}}+\dot{\beta}\frac{\partial L}{\partial \dot{Q}}=0
\end{eqnarray}
Using the equations (\ref{4.4}) and (\ref{4.17}), the following equations are obtained
\begin{eqnarray}
\alpha\frac{\partial L}{\partial a}=\alpha \Big[3\alpha^{3}(F-QF_{Q})-6F_{Q}\dot{a}^{2}-6\dot{a}^{2}\Big],
\end{eqnarray}
\begin{eqnarray}
\beta\frac{\partial L}{\partial Q}=-\beta\Big[a^{3}(QF_{QQ})+6F_{QQ}a\dot{a}^{2}\Big],
\end{eqnarray}
\begin{eqnarray}
\dot{\alpha}\frac{\partial L}{\partial \dot{a}}=-12a\dot{a}^{2}\frac{\partial \alpha}{\partial a}-12F_{Q}a\dot{a}^{2}\frac{\partial\alpha}{\partial a}-12a\dot{a}\dot{Q}\frac{\partial \alpha}{\partial Q}-12F_{Q}a\dot{a}\frac{\partial \alpha}{\partial Q},
\end{eqnarray}
\begin{eqnarray}
\dot{\beta}\frac{\partial L}{\partial \dot{Q}}=0.
\end{eqnarray}
Substituting equation (\ref{4.4}) into (\ref{4.17}) and using the relations $\dot{\alpha}=(\frac{\partial \alpha}{\partial a})\dot{a}+(\frac{\partial \alpha}{\partial Q})\dot{Q}$, $\dot{\beta}=(\frac{\partial \beta}{\partial a})\dot{a}+(\frac{\partial \beta}{\partial Q})\dot{Q}$.
As mentioned above, requiring the coefficients of $\dot{a}^{2}$, $\dot{Q}^{2}$ and $\dot{a}\dot{Q}$ to be zero, we find that
\begin{eqnarray}\label{4.22}
\alpha F_{Q}+ \alpha +\beta a F_{QQ}+2a \frac{\partial \alpha}{\partial a}+2F_{Q}a \frac{\partial \alpha}{\partial a}=0,
\end{eqnarray}
\begin{eqnarray}\label{4.23}
\alpha \frac{\partial \alpha}{\partial Q}=0,
\end{eqnarray}
\begin{eqnarray}\label{4.24}
3\alpha a^{2}(F-F_{Q}Q)-\beta a^{3}(F_{QQ}Q)=0.
\end{eqnarray}
The corresponding constant of motion (Noether charge) given by equation (\ref{eq:noether_charge}), reads
\begin{eqnarray}\label{4.25}
Q_{0}=-12\alpha F_{Q}a\dot{a}-12\alpha a\dot{a}=\rm Const.
\end{eqnarray}
A solution to the equations (\ref{4.22}), (\ref{4.23}), and (\ref{4.24}) is attainable if explicit expressions for $\alpha$ and $\beta$ are determined, and provided that at least one of these parameters is non-zero, thus the Noether symmetry condition is satisfied. Obviously, from equation (\ref{4.23}), it easy to see that $\alpha$ is independent of $Q$, and hence it is a function of $a$ only, {\it i.e}, $\alpha =\alpha (a)$.
On the other hand, from equation (\ref{4.24}), we have
\begin{eqnarray}
\beta =\frac{3\alpha(F-QF_{Q})}{aQF_{QQ}},
\end{eqnarray}
\begin{eqnarray}\label{4.28}
\alpha = \alpha_{0}a^{\frac{2QF_{Q}-Q-3F}{2Q+2QF_{Q}}},
\end{eqnarray}
considering the
\begin{eqnarray}
\frac{2QF_{Q}-Q-3F}{2Q+2QF_{Q}} =n,
\end{eqnarray}
\begin{eqnarray}\label{4.29}
2Q(n-1)F_{Q}+Q(2n+1)+3F=0,
\end{eqnarray}
by solving the differential equation (\ref{4.29}), we obtain
\begin{eqnarray}\label{4.30}
F(Q)=-Q+c(Q-n Q)^{\frac{3}{2-2n}},
\end{eqnarray}
where $c$ is an integration constant. To find the exact solution of $a(t)$ for this type of $f(Q)$, we can obtain an ordinary differential equation of $a(t)$, Thus, by using
\begin{eqnarray}\label{4.31}
c_{1}= c(1-n)^{\frac{3}{2-2n}},
\end{eqnarray}
\begin{eqnarray}\label{4.32}
c_{2}=\frac{3}{2-2n}c_{1},
\end{eqnarray}
\begin{eqnarray}\label{4.33}
c_{3}= \frac{2n+1}{2-2n}c_{2},
\end{eqnarray}
\begin{eqnarray}\label{4.34}
c_{4}=\Big[\frac{Q_{0}}{-12\alpha_{0}c_{2}(-6)^{\frac{2n+1}{2-2n}}}\Big]^{\frac{1}{2+n}},
\end{eqnarray}
and substituting equations (\ref{4.31}), (\ref{4.32}) and (\ref{4.33}) into equation (\ref{4.25}) and also applying equations (\ref{4.28}) and (\ref{4.30}), we obtain
\begin{eqnarray}\label{4.35}
\dot{a}a^{-n}=\Big[\frac{Q_{0}}{-12\alpha_{0}c_{2}(-6)^{\frac{2n+11}{2-2n}}}\Big]^{\frac{1-n}{2+n}},
\end{eqnarray}
from equation (\ref{4.35}), we get
\begin{eqnarray}
a(t)=(1-n)^{\frac{1}{1-n}}\Big[\frac{Q_{0}}{-12\alpha_{0}c_{2}(-6)^{\frac{2n+1}{2-2n}}}\Big]^{\frac{1}{2+n}}~~~t^{\frac{1}{1-n}}-(c_{5})^{\frac{1}{1-n}}(1-n)^{\frac{1}{1-n}}
\end{eqnarray}
where $c_{5}$ is an integration constant. Obviously, in the late time $|c_{4}t|\gg|c_{5}|$, and the universe experiences a power-law expansion.
In fact, we can make it clearer that by requiring $a(t=0)=0$, it is easy to see that the integral constant $c_{5}$ is zero. So, we have
\begin{eqnarray}\label{4.37}
a(t)\sim t^{\frac{1}{1-n}}.
\end{eqnarray}
It is important to say that the condition $n<1$ and $n\neq0$ is necessary to guarantee the expansion of the universe.


\section{Dynamical system approach}
\label{section5}
To analyze the dynamics of the $f(Q)$ model, we must assume that the conservation equation of the energy-momentum tensor holds for pressureless matter and radiation, respectively. Thus, we can write
\begin{eqnarray}\label{5.1}
\dot{\rho}_{m}+3H\rho_{m}=0,~~~~~~~~~~\dot{\rho}_{r}+4H\rho_{r}=0.
\end{eqnarray}
Using the following dimensionless variable
\begin{eqnarray}\label{5.2}
x=\frac{2QF_{Q}-F}{6H^{2}},~~~~~~~~~~~y=\frac{\rho_{r}}{6H^{2}}.
\end{eqnarray}
Other hand we have
\begin{eqnarray}\label{5.3}
6H^{2}=\rho_{m}+\rho_{r}+\rho_{de},~~~~~~~~~ \frac{\rho_{m}}{6H^{2}}+\frac{\rho_{r}}{6H^{2}}+\frac{2QF_{Q}-F}{6H^{2}}=1,
\end{eqnarray}
where $\rho_{m}=\rho_{m0}a^{-3}$, $\rho_{r}=\rho_{r0}a^{-4}$, the subscript $"0"$ indicates the present value of corresponding quantity.
Further, we can rewrite $\Omega_{r}=y$, $\Omega_{de}=x$, and $\Omega_{m}=1-x-y$,
where $\Omega_{r}$, $\Omega_{m}$, and $\Omega_{de}$ are the energy density parameters of the radiation, pressureless matter, and dark energy, respectively, in which these satisfy the relation
\begin{eqnarray}\label{5.4}
\Omega_{m}+\Omega_{r}+\Omega_{de}=1.
\end{eqnarray}
Taking the derivative of equation (\ref{5.3}) with respect to time $t$ yields
\begin{eqnarray}\label{5.5}
12H\dot{H}=\dot{\rho_{m}}+\dot{\rho_{r}}+\frac{d (2QF_{Q}-F)}{dt},
\end{eqnarray}
By substituting equations (\ref{5.1}) into equation (\ref{5.5}), we derive
\begin{eqnarray}
\dot{H}(1+2QF_{QQ}+F_{Q})=-\left(\frac{\rho_{m}}{4}+\frac{\rho_{r}}{3}\right).
\end{eqnarray}
Thus, we have
\begin{eqnarray}
\frac{\dot{H}}{H^{2}}=-\frac{1}{2}\frac{3-3x+y}{2QF_{QQ}+F_{Q}+1},
\end{eqnarray}
By using the functional form of $F(Q)$ from equation (\ref{4.30}), we have
\begin{eqnarray}
F_{Q}=-1+c_{2}Q^{\frac{2n+1}{2-2n}},
\end{eqnarray}
\begin{eqnarray}
F_{QQ}=c_{3}Q^{\frac{4n-1}{2-2n}}.
\end{eqnarray}
From equation (\ref{5.2}), we obtain the differential equations for $x$ and $y$,
\begin{eqnarray}
x'=\frac{-2\dot{H}}{H^{2}}\Big[x+(2QF_{QQ}+F_{Q})\Big],
\end{eqnarray}
\begin{eqnarray}
y'=-2y (2+\frac{\dot{H}}{H^{2}}),
\end{eqnarray}
where $N=\ln a$, and prime denotes a derivative with respect to $\ln a$.\\
By taking
\begin{eqnarray}
\beta=\frac{3}{2-2n},
\end{eqnarray}
we can write
\begin{eqnarray}
2QF_{QQ}+F_{Q}=\beta(1-x)-1.
\end{eqnarray}
Therefore, by using the above equation we can get
\begin{eqnarray}\label{5.14}
x'=\frac{1}{\beta}[\beta(3-3x+y)-3(1-x)-y],
\end{eqnarray}
\begin{eqnarray}\label{5.15}
y'=\frac{1}{\beta}[y(1-x)^{-1}-4\beta+3].
\end{eqnarray}
In the following, we find it useful to define an effective of state $\omega_{eff}$, which is related to the Hubble parameter as $\frac{\dot{H}}{H^{2}}=(-\frac{3}{2})(1+\omega_{eff})$.
Using the above equations we find
\begin{eqnarray}\label{5.16}
\omega_{eff}=\frac{1-x-\beta(1-x)+\frac{y}{3}}{\beta(1-x)},
\end{eqnarray}
where $\omega_{eff}$ is effective equation of state.

\subsection{The Jacobian matrix and stability analysis}

To analyze the cosmological dynamics of the model, we extract the critical points 
$(\ref{5.14})$ and $(\ref{5.15})$ by solving the equations $x'=0$ and $y'=0$.
We construct the Jacobian matrix and consider the stability points.\\
Consider the autonomous system with the dynamical variables $(x, y)$:
\begin{eqnarray}
x'=f_{1}(x,y),~~~~~y'=f_{2}(x,y),
\end{eqnarray}
\begin{eqnarray}
f_{1}(x,y)=\frac{2(1-n)}{3}\Big[\beta(3-3x+y)-3(1-x)-y\Big],
\end{eqnarray}
\begin{eqnarray}
f_{2}(x,y)=\frac{2(1-n)}{3}\Big[y(1-x)^{-1}-4\beta+3\Big].\\
\end{eqnarray}
\[
J=\begin{bmatrix}
-3\beta+3 & \beta-1 \\
y & 1+x
\end{bmatrix}
\]
\\
The eigenvalues of the Jacobian matrix are obtained from the following equation,
\begin{eqnarray}
\lambda=\frac{(-3\beta+ x+ 4)\pm\sqrt{[(-3\beta+x+4)^{2}+4(\beta-1)(3+3x+y)]}}{2},
\end{eqnarray}
The coordinates of the fixed points of the dynamical system are shown in Table 1.
\begin{table}[H]
\centering
\caption{Fixed points and their stability}
\begin{tabular}{|c|c|c|c|c|c|c|}
\hline
Label & $x$ & $y$ & $ W_{\rm eff}$  & Eigenvalues & $n$ & Stability \\
\hline
A & $ 1.000 $ & $0.000$ & $\infty$ & $1.750 , -6.000$& Arbitrary& Saddle \\
\hline
B & $0.000$ & $1.000$ & $\frac{1}{3}$ & $1.750\pm\sqrt{\frac{-15}{16}}$& -1& Oscillatory instability\\
\hline
C & $0.000$ & $0.000$ & $ 0.000 $  & $1.000, 0.000$& -0.5 & Unstable \\
\hline
D & $0.750$ & $0.000$ & $-\frac{2}{3}$ & $1.250,-6.250$& 0.5& Sadlle\\
\hline
\end{tabular}
\end{table}
Here, we highlighted some critical points as follows\\
Case I: $(x=0.75 ~~ y=0~~\Omega_{m}=0.25, \Omega_{de}=0.75~~ n=0.5)$, we have $\omega_{eff}=-\frac{2}{3}$,~~~$a(t)\sim t^{2}$.
It shows $\Lambda$CDM model for universe.\\
Case II: $~(x=0~~y=0~~\Omega_{m}=1~~n=-0.5)$,~~we have $\omega_{eff}=0~~a(t)\sim t^{\frac{2}{3}}$.
It shows matter-dominated universe model.
Case III:~$(x=0~~y=1~~\Omega_{m}=0~~n=-1)$,~~we have $\omega_{eff}\sim\frac{1}{3}~~a\sim t^{\frac{1}{2}}$.
It shows radiation-dominated  universe model.\\
Case IV:~$(x=1~~y=0~~\omega_{eff}=\infty~~P>\frac{1}{3}\rho)$.
It shows the singularity of the universe in the future.
\\
To understand general behavior of equations (\ref{5.14}) and (\ref{5.15}), in  figure 1, we plot the trajectories of solutions in the $(x, y)$ plane for $n = 0.5,-1,-0.5$
\begin{figure*}[ht]
  \centering
  \includegraphics[width=2in]{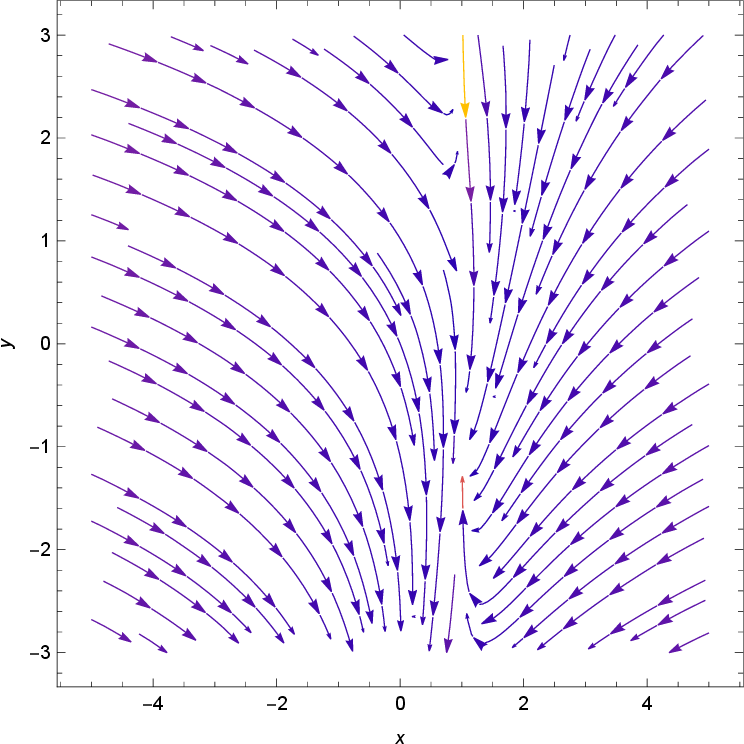}\hspace{1cm}
  \includegraphics[width=2in]{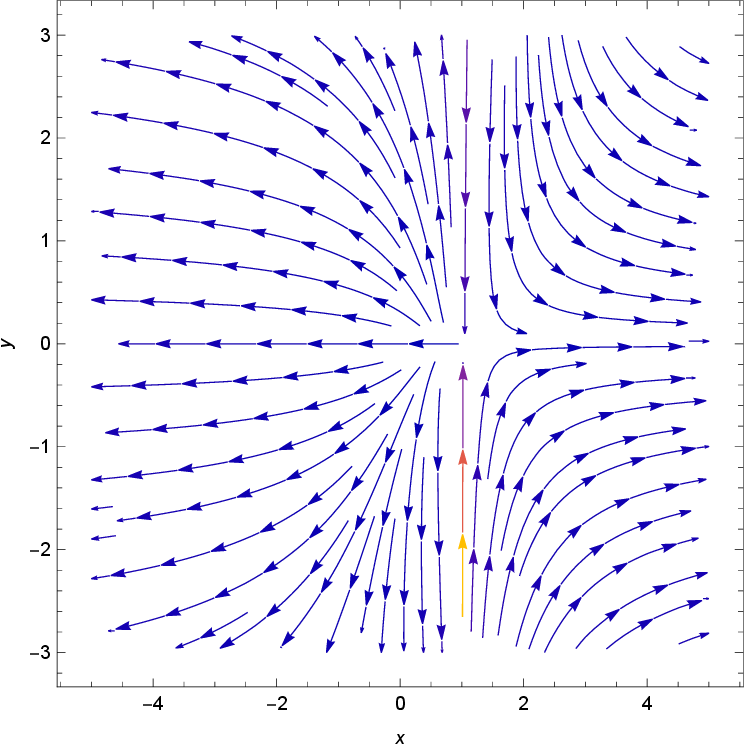}\hspace{1cm}
  \includegraphics[width=2in]{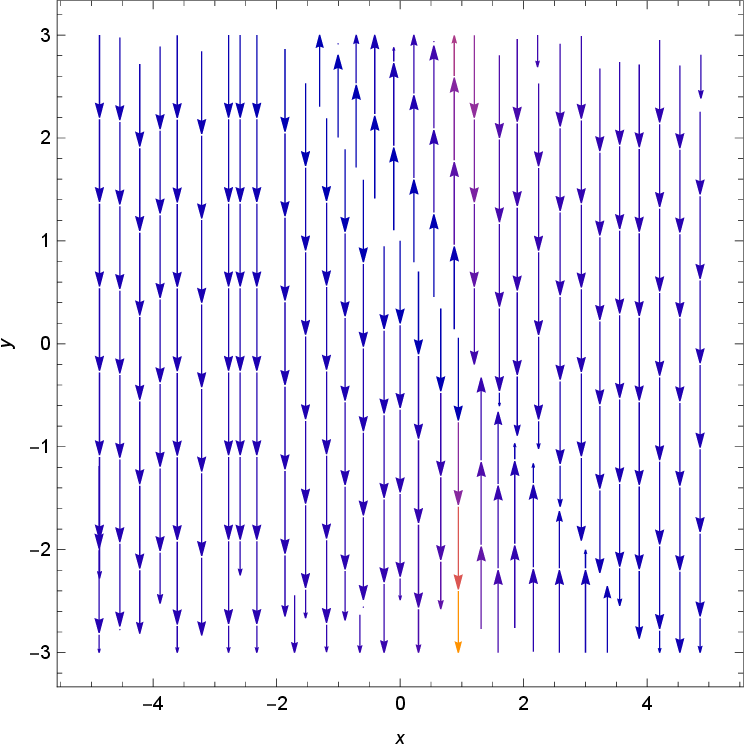}\hspace{1cm}
    \caption{A phase portrait of $\Lambda$CDM, radiation-dominated, and matter-dominated universe from left to right, respectively.}
  \label{stable}
\end{figure*}

\subsection {Examining the solutions with observational data in the inflationary regime}

In this subsection, we take a glance at the observational data by applying the solutions obtained in Section IV to describe an inflationary universe.\\
The slow-roll parameters can be defined as follows:
\begin{eqnarray}
\varepsilon_{1}=-\frac{\dot{H}}{H^{2}},  ~~~~~~~~~~~\varepsilon_{n+1}=\frac{\dot{\varepsilon_{n}}}{H\varepsilon_{n}},
\end{eqnarray}
where $ (n=1, 2, 3,....) $.
Accordingly, the scalar spectral index $n_{s}$ and the tensor-to-scalar ratio
$r$ can be expressed as
\begin{eqnarray}\label{5.23}
n_{s}=[1-2\varepsilon_{1}-2\varepsilon_{2}]\mid_{h.c.},~~~~~~~~~ r=16\varepsilon_{1}\mid_{h.c.},
\end{eqnarray}
where the subscript $`h.c`$ stands for the horizon crossing. Based on the 2018 Planck results, the observational indices are constrained as follows
\begin{eqnarray}
~~~~~~~n_{s}=0.9649\pm 0.0042 , ~~~~~~~~ r< 0.064.
\end{eqnarray}
For power-law solutions, the Hubble parameter takes the following form $H \sim \frac{m}{t}$.
\\
On the other hand, according to equation (\ref{4.37}), we have $ a(t)\sim t^{\frac{1}{1-n}}$, thus
by considering $m\equiv \frac{1}{1-n} $, we can write
\begin{eqnarray}
\varepsilon_{1}=\frac{1}{m},~~~~~~~~~~\varepsilon_{1}=1-n.
\end{eqnarray}
Because of $\varepsilon_{1}>0$, we have $n<1$.\\
According to the tensor-to-scalar ratio $r$, we insist the power of $t$ in the scale factor, namely $n$, must be consistence with data, thus the bound on $n$ is given by
\begin{eqnarray}
~~~ 16(1-n)<0.0649,~~~~~~n<0.9959,~~~~~~\varepsilon_{1}<0.0040.
\end{eqnarray}
On the other hand, by definition
\begin{eqnarray}
\varepsilon_{2}=\frac{\dot{\varepsilon_{1}}}{H\varepsilon_{1}},~~~~~~\dot{\varepsilon_{1}}=-\frac{\ddot{H}}{H^{2}}+2\frac{\dot{H^{2}}}{H^{3}},\\
 ~~~~~~~~~ \varepsilon_{2}=\frac{[\frac{-2}{m t }+\frac{2}{m t}]}{t^{-1}}=0.
\end{eqnarray}
Substituting the results into equation (\ref{5.23}), we obtain
\begin{eqnarray}~~~~n_{s}\simeq\Big[1-0.008-0\Big]\simeq 0.9920\pm 0.0042.
\end{eqnarray}
The connection between theoretical results and observational data in an inflationary universe is given by these parameters.


\section{ $f(Q)$ scalar-tensor cosmology via Noether symmetry}
Let us consider the general action
\begin{eqnarray}\label{6.1}
S=\int d^{4}x  \sqrt{-g}(\phi^{2}f(Q)+4\omega(\phi)g^{\mu\nu}\nabla_{\mu}\phi\nabla_{\nu}\phi- V(\phi)).
\end{eqnarray}
where the scalar field  $\phi$  is  non-minimally coupled to $f(Q)$, and $\omega(\phi)$ and $ V(\phi)$  are respectively the in Brans-Dicke parameter and potential as generic functions of $\phi$.
One can define a canonical Lagrangian $L =L (a, \dot{a}, Q, \dot{Q}, \phi, \dot{\phi})$, where dot denotes derivative with respect to the cosmic time $t$. The variable $a$ is scale factor in FRW universe, the configuration space $\emph{T}\emph{Q}=({a, \dot{a}, Q, \dot{Q}, \phi, \dot{\phi}})$, and all dynamical variables $a$, $Q$ and $\phi$ are assumed to depend just on $t$ to restor in the homogeneity and isotropy.
It is assumed that $Q$, $a$ and $\phi$, are canonical variables.
Thus, one can use the method of Lagrange multipliers in this theory as
\begin{eqnarray}\label{6.2}
S=\int dt a^{3} \{\phi^{2}f(Q)+4\phi^{2}\omega(\phi)-V(\phi)+\zeta\left[Q+6\left(\frac{\dot{a}}{a}\right)^{2}\right]\},
\end{eqnarray}
where $\zeta$ is a Lagrange multiplier.
The variation of action with respect to $Q$ gives $\zeta=-\phi^{2}f_{Q}$, where $f_{Q}=\frac{df}{dQ}$.
Therefore, the action (\ref{6.2}) can be rewritten as
\begin{eqnarray}\label{6.3}
S=\int dt a^{3}\{\phi^{2}f(Q)+4\phi^{2}\omega(\phi)- V(\phi)-\phi^{2}f_{Q}\left[Q+6\left(\frac{\dot{a}^{2}}{a^{2}}\right)\right]\}.
\end{eqnarray}
By integrating by parts, we obtain the point-Like FRW Lagrangian as
\begin{eqnarray}\label{6.4}
L= a^{3}\phi^{2}(f(Q)-f_{Q}) +6 \phi^{2}a \dot{a}^{2}f_{Q}+a^{3}(4\dot{\phi}^{2}\omega(\phi)-V(\phi)),
\end{eqnarray}
By using  equations (\ref{eq:vector_field}) and (\ref{eq:lie_derivative}), we have
\begin{align}\label{6.5}
XL&=3A a^2 \phi^2 (f(Q)-Qf_{Q})+6A\phi^2\dot{a}^2 f_{Q}+12 A a^2\dot{\phi}^2 \omega -3A a^2V(\phi)-Ba^3 \phi^2 Q f_{QQ}
+\nn&~~~~~ 6B\phi^2 a \dot{a}^2 f_{QQ}+2C a^3 \phi(f(Q)-Qf_{Q})+12C\phi a\dot{a}^2 f_{Q}+4C a^3 \dot{\phi}^2 \frac{d \omega }{d \phi }-C a^3\frac{d V}{d \phi}+
12\phi^2 a \dot{a}^2 f_{Q}\frac{\partial A}{\partial a}+\nn&~~~~~12 \phi^2 a \dot{a}\dot{Q}f_{Q}\frac{\partial A}{\partial Q}+
12\phi^2 a \dot{a}\dot{\phi}f_{Q}\frac{\partial A}{\partial \phi}+ 8a^3 \dot{\phi} \dot{a} \omega \frac{\partial C}{\partial a}+
8a^3\dot{\phi} \dot{Q}\omega\frac{\partial C}{\partial Q}+8a^3\dot{\phi}^2 \omega \frac{\partial C}{\partial \phi}=0.
\end{align}
By setting to zero the coefficients of the terms $\dot{a}^{2}$, $\dot{Q}^{2}$, $\dot{\phi}^{2}$, $\dot{a}\dot{Q}$, $\dot{a}\dot{\phi}$ and $\dot{\phi}\dot{Q}$ in ${\cal L}_{X}L=0$.
Finally, we have obtain
\begin{eqnarray}\label{6.6}
A \phi f_{Q}+B \phi a f_{QQ}+2C a f_{Q}+2\phi a f_{Q} (\frac{\partial A}{\partial a})=0,
\end{eqnarray}
\begin{eqnarray}\label{6.7}
3A\omega+C a \frac{d \omega}{d \phi}+ 2 a \omega \frac{\partial C}{\partial \phi}= 0,
\end{eqnarray}
\begin{eqnarray}\label{6.8}
 12\phi^2 a f_{Q}\frac{\partial  A }{\partial Q}=0,~~~~~~ \frac{\partial A}{\partial Q}=0,
\end{eqnarray}
\begin{eqnarray}\label{6.9}
3\phi^2 f_{Q}\frac{\partial A}{\partial \phi}+2a^2\omega \frac{\partial C}{\partial a}=0,
\end{eqnarray}
\begin{eqnarray}\label{6.10}
8a^3 \omega \frac{\partial C}{\partial Q }=0,~~~~~~~~~\frac{\partial  C}{\partial Q }=0,
\end{eqnarray}
\begin{eqnarray}\label{6.11}
3A \phi^{2} (f(Q)-Qf_{Q})-3A  V(\phi)-B a\phi^2 Q f_{QQ}+2C a\phi(f(Q)-Qf_{Q})-C a \frac{d V}{d \phi}=0.
\end{eqnarray}
A solution of equations (\ref{6.6})-(\ref{6.11}) exist if explicit form of $A$, $B$ and $C$ are found.

The Friedmann equations is obtained by construction of the zero Hamiltonian constraint as
\begin{eqnarray}\label{6.12}
{\cal H}=\frac{\partial L}{\partial q\dot{_{i}}}\dot{q_{i}}-L=\dot{a}\frac{\partial L}{\partial \dot{a}}+\dot{Q}\frac{\partial L}{\partial \dot{Q}}+\dot{\phi}\frac{\partial L}{\partial \dot{\phi}}-L,
\end{eqnarray}
\begin{eqnarray}\label{6.13}
\phi^{2}Qf_{Q}-4\dot{\phi^{2}}\omega+\phi^{2}(f-Qf_{Q})-V(\phi)=0,
\end{eqnarray}
substituting equation (\ref{6.13}) into equation (\ref{6.5}) we obtain
\begin{eqnarray}\label{6.14}
A\phi f_{Q}+B\phi a f_{QQ}+2ca f_{Q}+2\phi a f_{Q}\frac{\partial A}{\partial a}=0,
\end{eqnarray}
\begin{eqnarray}\label{6.15}
6A\omega +2Ca\frac{d \omega}{d \phi}+2a\omega \frac{\partial C}{\partial \phi}=0,
\end{eqnarray}
\begin{eqnarray}
\frac{\partial A}{\partial Q}=0,
\end{eqnarray}
\begin{eqnarray}
2a^{2}\omega \frac{\partial C}{\partial a}+3\phi^{2}f_{Q}\frac{\partial A}{\partial \phi}=0,
\end{eqnarray}
\begin{eqnarray}
\frac{\partial C}{\partial Q}=0,
\end{eqnarray}
\begin{eqnarray}
-3A\phi f_{Q}-Ba\phi f_{QQ}-2Caf_{Q}=0,
\end{eqnarray}
then we have
\begin{eqnarray}
B=-\left[\frac{A}{a}+\frac{2C}{\phi}+2\frac{\partial A}{\partial a}\right]\frac{f_{Q}}{f_{QQ}}.
\end{eqnarray}
\begin{eqnarray}
A=a\frac{\partial A}{\partial a}.
\end{eqnarray}
\begin{eqnarray}
C=\frac{1}{k^{2}}\left[3\frac{\partial A}{\partial a}+\frac{\partial C}{\partial \phi}\right].
\end{eqnarray}
\begin{eqnarray}\label{6.23}
k^{2}=\frac{1}{c}\left[3\frac{A}{a}+\frac{\partial C}{\partial \phi}\right].
\end{eqnarray}
substituting equation (\ref{6.23}) into equation (\ref{6.7}) then we obtain
\begin{eqnarray}
\frac{d \omega}{d \phi}+k^{2}\omega=0,
\end{eqnarray}
where $k^2>0$, we have the oscillating solutions
\begin{eqnarray}\omega(\phi)=\omega_{0} \exp (\pm ik\phi),
\end{eqnarray}
where as for the choice leading to $k^{2}<0$, we obtain exponential solutions as
\begin{eqnarray}
\omega(\phi)=\omega_{0} \exp (\pm k\phi).
\end{eqnarray}
Therefor, we find
\begin{eqnarray}
A=c_{1}g_{0}a,~~ C=C_{0}, ~~ B=-\left[3c_{1}g_{0}+\frac{2C}{\phi}\right]\frac{f_{Q}}{f_{QQ}},
\end{eqnarray}
where ~~ $C(a,\phi)=C_{0}$~ and ~ $g(\phi)=g_{0}$.
We may find the constant of motion,namely the Noether charge as
\begin{eqnarray}
\Theta_{0}=A\frac{\partial L}{\partial \dot{a}}+B\frac{\partial L}{\partial \dot{\phi}}+C\frac{\partial L}{\partial \dot{Q}},
\end{eqnarray}
\begin{eqnarray}
\Theta_{0}=-6\phi a^{2}c_{1}g_{0}\left[2f_{Q}\dot{(\phi a)} \right]-12C_{0}a^{2}\dot{a}\phi f_{Q}+8C_{0}a^{3}\dot{\phi}\omega(\phi)+6\phi^{2}a^{2}\dot{a}\left[\frac{2C_{0}}{\phi}+3c_{1}g_{0}\right]f_{Q}.
\end{eqnarray}
This equation may be written as
\begin{eqnarray}f_{QQ}\dot{Q}=-\frac{\Theta_{0}}{6a^{3}\phi^{2}c_{1}g_{0}}-2\frac{d}{d t}\ln(a\phi)f_{Q}+\frac{4C_{0}\dot{\phi}\omega(\phi)}{3\phi^{2}c_{1}g_{0}}+3f_{Q}H.
\end{eqnarray}
The Hamiltonian constraint from equation (\ref{6.12}) is given by
\begin{eqnarray}
{\cal H}=f(Q) +6f_{Q}H^{2}+12f_{Q}H\left(\frac{\dot{\phi}}{\phi}\right)-4(\frac{\dot{\phi}}{\phi})^{2}\omega(\phi)-f_{Q}Q -\frac{V(\phi)}{\phi^{2}}=0,
\end{eqnarray}
or it can be written as
\begin{eqnarray}
{\cal H}=f(Q)- \frac{\Theta H}{\phi^{2}a^{3}c_{1}g_{0}}+12f_{Q}H^{2}-f_{Q}Q-\frac{V(\phi)}{\phi^{2}}+
\frac{\dot{\phi}}{\phi^{2}}\omega(\phi)\left(\frac{8HC_{0}}{c_{1}g_{0}}-4\dot{\phi}\right)=0.
\end{eqnarray}
The Friedmann equation in this model can be obtained from the zero Hamiltonian constraint(${\cal H}=0$).

\section{Conclusion}

We have studied  $f(Q)$ gravity, via Noether symmetry and by utilizing this symmetry within the framework of $f(Q)$ gravity to derive the functional expression for $f(Q)$, which is given by $f(Q)=c(Q-nQ)^{\frac{3}{2-2n}}$.
To confirm the exact solution of the model through Noether symmetry, we have considered the FRW universe with the dynamic solution of the system using dimensionless variables. We have shown that the accelerated expansion of the universe follows a power law scale factor as $a(t)\sim t^{\frac{1}{1-n}}$. For the values of $n<1$, the universe enters the accelerated expansion phase.
Also in the continuation, we have obtained the quantities corresponding to the exact solution, and again for $n<1$, it will be accelerated according to the universe's expansion. We have discussed the cosmological dynamics of the system using the dimensionless quantities, and by changing the values of $n$, the matter-dominated and the $\Lambda$CDM universe can be obtained.
In section VI, within the context of $f(Q)$ scalar-tensor cosmology, we have utilized the Noether symmetry method to identify cosmological models that align with the Noether symmetry theorem.

\section*{Data Availability Statement}
 No data associated in the manuscript.

\end{document}